\documentstyle[12pt,epsfig]{article}
\begin{document}
\bibliographystyle{unsrt}  
\newcommand{\npb}[3]{Nucl.~Phys.~B #1 (#2) #3}
\newcommand{\plb}[3]{Phys.~Lett.~B #1 (#2) #3}
\newcommand{\prd}[3]{Phys.~Rev.~D #1 (#2) #3}
\newcommand{\prl}[3]{Phys.~Rev.~Lett. #1 (#2) #3}
\newcommand{\zpc}[3]{Z.~Phys.~C #1 (#2) #3}
\newcommand{\cpc}[3]{Comput.~Phys.~Commun. #1 (#2) #3}
\rightline{CERN/TH 98-238}
\rightline{FNT/T 98/07}
\begin{center}
{\Large\bf Single- and multi-photon final states with \\ 
missing energy at $e^+ e^-$ colliders}
\end{center}
\vskip 24pt
\begin{center}
{\large
G.~Montagna$^a$, M.~Moretti$^{b,c}$, O.~Nicrosini$^a$ and F.~Piccinini$^a$}
\end{center}
\vskip 24pt
\begin{center}
$^a$ Dipartimento di Fisica Nucleare e Teorica 
- Universit\`a 
di Pavia, and \\
INFN - Sezione di Pavia, Via A. Bassi 6, Pavia, Italy\\
\vskip 12pt\noindent
$^b$ Theory Division, CERN,  CH-1211 Geneva 23, 
Switzerland\\
\vskip 12pt\noindent
$^c$ Dipartimento di Fisica - Universit\`a di    
Ferrara, and \\
INFN - Sezione di Ferrara, Ferrara, Italy
\end{center}
\vskip 48pt
\begin{abstract}
The search for new physics in single- and multi-photon 
final states with 
large missing energy at LEP and future $e^+ e^-$ colliders requires 
precise predictions for the Standard Model irreducible background.   
While at LEP1 the theoretical situation is under control, 
going to LEP2  
(and beyond) some improvements are necessary. To approach the aimed 
$O(1\%)$ theoretical accuracy, the tree-level matrix elements 
for the processes $e^+ e^- \to \nu \bar\nu n\gamma$, with $n=1,2,3$, 
are exactly computed in the Standard Model, including the possibility 
of anomalous couplings for single-photon production. 
Due to the presence of observed photons in the final state, 
particular attention is paid to the treatment of higher-order 
QED corrections. Comparisons with existing calculations 
are shown and commented. An improved version of the event generator 
NUNUGPV  is presented.

\end{abstract}

\section{Introduction}

The production of one or more photons and missing energy 
in high-energy 
electron-positron ($e^+ e^-$) collisions is a process of great 
interest for 
the scientific programme of LEP and future $e^+ e^-$ linear 
colliders~\cite{expref,exp98}.

In the context of precision tests of the electroweak interactions 
on top of the $Z$ resonance (LEP1)~\cite{yr86,yr89}, 
the Standard Model (SM) 
process $e^+ e^- \to \nu \bar\nu \gamma$, where $\nu$ refers to 
light neutrinos 
and $\gamma$ to a detected, energetic photon, has been 
successfully used for the determination of the number of light 
neutrino 
species, in agreement with the result obtained via 
the measurement of the partial width  
of the $Z$-boson into invisible particles.

Above the $Z$ resonance, i.e. for current data taking in the LEP2 
energy range ($\sqrt{s} \simeq 160$-$200$~GeV)~\cite{yr96}
 and for planned 
experiments at the TeV scale~\cite{lc}, the events with single- and 
multi-photon final states plus missing energy ($\rlap{$\, \, /$}E$) 
play an important role in the search for new phenomena 
beyond the SM~\cite{expref,exp98}. Actually, the SM processes 
$e^+ e^- \to \nu \bar\nu n \gamma$, with $n=1,2,\dots$, are the 
largely dominating irreducible backgrounds 
to a New Physics (NP) signature 
consisting of one or more photon(s) and nothing 
else seen in the detector. 
Such events can indeed originate from various mechanisms, both in 
gravity- and gauge-mediated supersymmetric models~\cite{newphys} 
as well 
as in scenarios with strong electroweak 
symmetry breaking~\cite{newphyss}. 
In supersymmetric extensions of the SM, neutralinos, gravitinos 
and sneutrinos 
are the lightest supersymmetric particles (LSP) yielding the 
content of 
missing energy to the events. Another interesting example of new 
phenomenon giving rise to $n \gamma + \rlap{$\, \, /$}E$ final states 
is the production of a pair of fourth-generation, heavy neutrinos 
in association with initial-state radiation (ISR). 
Finally this signature can be useful to study anomalous 
triple and quartic gauge couplings. In particular, it can be 
conveniently used for the study of anomalous couplings as a 
complementary channel with respect to  processes such as 
$e^+ e^- \to 4f,6f$, without contamination 
coming from the vertices involving at least three massive 
gauge bosons.

The present situation of the theoretical calculations 
for the above quoted
processes can be considered as satisfactory for the purposes of data 
analysis at LEP1. Going to LEP2, the typical SM cross section 
is of the order
 of a few picobarn, yielding thousands of events collected 
in the four LEP experiments. Hence, there is a demand for 
theoretical predictions with an accuracy 
 of the order of 1\% for the rate of $e^+ e^- \to
 \nu \bar\nu n\gamma$ events in the SM. Furthermore, as it will 
 be discussed in the following, the situation 
concerning the theoretical  calculations is not 
completely satisfactory. 
In particular, a careful treatment of higher-order QED corrections 
to processes with detected photons in the final 
state becomes mandatory 
for a meaningful comparison between data and theory. 

Given the above physics motivations, the aim of the present paper is 
to perform an {\em exact} tree-level 
calculation in the SM of the $e^+ e^- \to \nu \bar\nu n \gamma$ cross 
sections, with $n=1,2,3$, supplemented with the 
most phenomenologically 
relevant and presently under control radiative corrections. 
While for one and two photons in the final state exact matrix element
calculations have recently appeared in the literature, as it will 
be discussed in detail in the next Section, no results for three
photons are available yet. Concerning ISR, 
particular care is devoted to the implementation of the effects due to
(undetected) photon emission.
An improved version of the event generator NUNUGPV~\cite{nunugpv}, 
based on the results here described, is presented for data analysis.  

The content of the paper is as follows. In Sect.~2 
the status of the available calculations and related programs is 
critically reviewed, putting special emphasis on the
 needs of improvement 
for current experiments at LEP. In Sects.~3 and 4 it is 
described how some of the existing approximations are overcome in 
the present study, providing various numerical results. 
The exact SM calculation of the lowest-order 
amplitudes for the signatures $e^+ e^- \to 
\nu \bar\nu n\gamma$, with $n=1,2,3$, is discussed in Sect.~3, 
while Sect.~4 is devoted to the implementation of higher-order 
QED radiative corrections. In passing, numerical 
results for integrated cross sections as well as comparisons with 
existing calculations are shown and commented. 
The impact of the theoretical 
improvements on exclusive photon distributions is 
discussed and compared with typical NP effects 
in Sect.~5. The main conclusions and open issues 
are drawn in Sect.~6. 

\section{Status of the theoretical predictions and generators}

\begin{figure}[h]
\begin{center}
\epsfig{file=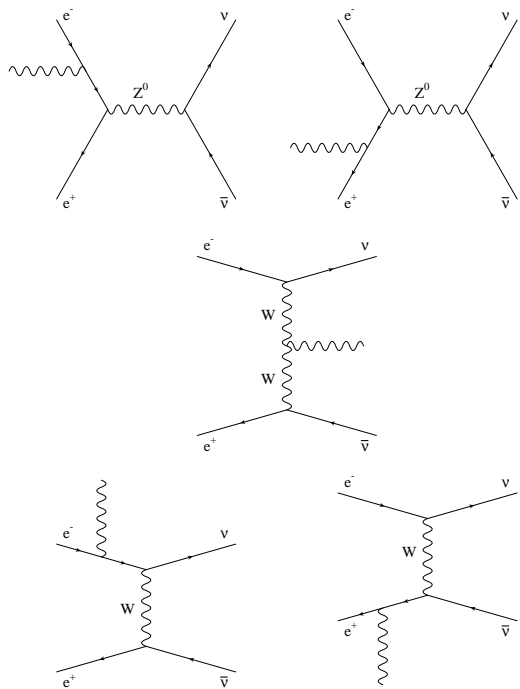,height=9truecm}
\end{center}
\caption{The tree-level Feynman diagrams for the process
$e^+ e ^- \to \nu \bar \nu \gamma$.}
\label{fig:fig1}
\end{figure}

Concerning the process $e^+ e^- \to \nu \bar\nu \gamma$, 
described by the  
Feynman diagrams depicted in Fig.~\ref{fig:fig1}, 
several calculations, 
with a different degree of accuracy, are known in the literature. 
Before the start of LEP/SLC operations, the status of the 
theoretical calculations 
and related computational tools has been summarized 
in Ref.~\cite{yr89nc} and, 
more recently, in view of the energy upgrade of the LEP collider from 
LEP1 to LEP2 regime, in Ref.~\cite{yr96sm}.

First attempts~\cite{pia} to compute the cross section associated to 
the graphs of Fig.~\ref{fig:fig1} consider the limit 
$M_W \to \infty$ and 
neglect the contribution of the non-abelian $WW\gamma$ 
vertex (the so-called 
Point Interaction Approximation). In other approximate 
calculations the 
invisible neutrino-pair cross section is dressed with
 some (universal) 
radiation factor to attach one external photon 
to the charged fermion legs, 
e.g. by using an angular dependent radiator~\cite{nt89,mnpt95}, 
a parton 
shower (PS) algorithm (as in the program PYTHIA)~\cite{pythia} 
or the 
Yennie-Frautschi-Suura (YFS) exclusive exponentiation (as done 
in KORALZ)~\cite{koralz}. By construction, these approaches allow 
to account for the leading contributions due to the collinear and 
infrared singularities but need to be corrected for the effect of 
sub-leading terms and/or internal photon radiation from the 
off-shell $W$ boson   
(as recently done in an approximate way 
in Ref.~\cite{koralnew}) that are contained in the exact 
matrix element. The first complete calculation of the matrix element 
of the process $e^+ e ^- \to \nu \bar \nu \gamma$ 
was done in Ref.~\cite{bbmmn88}. The corresponding 
exact matrix element is 
implemented in the event generator MMM~\cite{mmm}. 
In Ref.~\cite{bbmmn88}, by working in the 
approximation of neglecting in the squared matrix element 
terms with at least 
three boson propagators, also a compact, analytical expression 
for the differential spectrum on the energy and angle of the 
observed photon is obtained, yielding the result 
\begin{equation}
{{d\sigma}\over {d \cos \vartheta_\gamma d k}} = 
{{\alpha}\over {12 \pi^2}}
G_F^2 M_W^4 {{s' k}\over {s k_+ k_-}}[\eta_+^2 F(\eta_+) + 
\eta_-^2 F(\eta_-)] .
\label{eq:bs}
\end{equation}
The meaning and explicit expression of the symbols entering 
eq.~(\ref{eq:bs}) can be found in Ref.~\cite{bbmmn88}. 
The photon spectrum of eq.~(\ref{eq:bs}) contains the 
bulk of the contributions due to $W$-boson exchange and 
agrees within 1\% with   
approximate calculations discussed above for center of mass 
(c.m.) energies 
around the $Z$ resonance~\cite{mnpt95}. The analytical photon 
spectrum of eq.~(\ref{eq:bs}) 
is implemented in the event generator NUNUGPV~\cite{nunugpv}.

\begin{figure}[ht]
\begin{center}
\epsfig{file=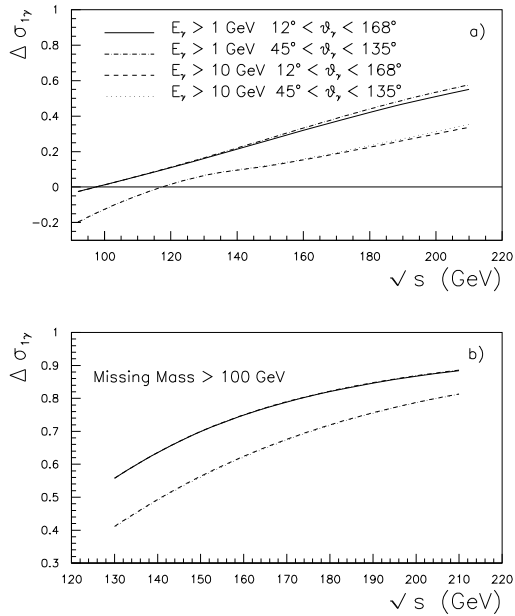,height=9truecm}
\end{center}
\caption{The relative size of the $W$-exchange 
diagrams with respect to the total cross section of the  
signature $\nu \bar \nu \gamma$, for different experimental cuts. 
The quantity 
$\Delta \sigma_{1\gamma}$ is defined as: 
$\Delta \sigma_{1\gamma} = 1 - 
3\sigma(\nu_{\mu} {\bar \nu}_{\mu} \gamma) / 
\sigma(\nu \bar \nu\gamma )$.
In Fig.~2a the four lines correspond to the 
cuts on the photon energy and angle specified in the plot; 
Fig.~2b is the same 
as Fig.~2a, with an additional cut on the 
missing mass of the event.}
\label{fig:fig2}
\end{figure}

Concerning radiative corrections, the exact one-loop 
electroweak corrections 
to $e^+ e ^- \to \nu \bar \nu \gamma$ process are not yet available. 
What is 
presently known is the full set of one-loop electroweak corrections to 
the ``sub-process'' $e^+ e ^- \to Z\gamma$, 
with an on-shell final $Z$ boson~\cite{bs87}, and the subset of 
one-loop QED corrections to the $Z$-exchange contribution to 
$\nu \bar\nu \gamma$ 
final state~\cite{bbmmn88,olqed}. However, since, contrary to LEP1, 
the $W$-boson contributions are essential at LEP2, as clearly 
shown in Fig.~\ref{fig:fig2} for different selection criteria, 
the still missing part of the full one-loop calculation due 
to the corrections to the dominant diagrams with $W$-boson exchange 
is necessary in order to reach the aimed theoretical 
precision at the 1\% level. 
Actually, one can see from 
Fig.~\ref{fig:fig2} that the relative contribution of $W$-diagrams
(squared modulus and interference terms) can amount 
to about 60\%(80\%) 
of the full cross section, including or excluding, via 
a cut on the missing mass of the event, the $Z$ return.
Therefore, in the absence of a full $O (\alpha)$ 
electroweak calculation, the goal of a 1\% theoretical accuracy in the 
predictions for single-photon plus missing energy production 
turns out to be   
difficultly reachable at present. However, in order to take care of 
the most sizeable higher-order corrections, the lowest-order 
calculations 
are typically improved by the inclusion of the (large) effects due 
to ISR. In mostly used computational tools such 
a contribution is taken 
into account via traditional algorithms for computing QED 
radiative corrections in the leading logarithmic (LL) approximation, 
such as the PS algorithm~\cite{ps} (as in PYTHIA), the Structure 
Function (SF) approach~\cite{sfcoll} (as in MMM and NUNUGPV, 
but also in PYTHIA) and YFS exclusive 
exponentiation~\cite{yfs} (as in KORALZ). Two different variations of 
the SF method are employed in MMM and NUNUGPV. 
SFs in strictly collinear 
approximation are used in MMM, while in NUNUGPV $p_t$-dependent 
SFs~\cite{nt89,nt89plb} are implemented to improve the treatment of 
ISR with  
the effect of the transverse degrees of freedom at the 
LL level~\cite{nunugpv} (this point will be further discussed later). 

The previously quoted programs KORALZ, MMM and NUNUGPV are the 
standard Monte Carlo generators used by the LEP collaborations for
 the analysis 
of the data relative to the events 
$e^+ e^- \to \nu \bar\nu \gamma (\gamma)$~\cite{exp98}. 
As already 
emphasized, these programs differ both in the treatment 
of the lowest-order 
matrix element and in the implementation of higher-order corrections. 
Actually, the agreement between the above generators, 
whenever compared to estimate the theoretical precision 
against the statical 
accuracy in the measured cross section 
(of the order of a few per cent), 
is fairly good for exclusive single-photon final states (say for 
photon energies above approximatively 10 GeV), whereas for 
low photon energies 
and more inclusive final states the agreement is not at present 
satisfactory~\cite{bus}. Therefore, since in the search
 for new stable, 
neutral particles one is interested in detecting an excess of events 
at low photon energies~\cite{bus}, the present status points out the 
need of improving theoretical predictions to avoid a 
loss of sensitivity 
to NP searches in radiative events at LEP2.

Let us come now to discuss the present status of the theoretical 
predictions  
for the process $e^+ e^- \to \nu \bar\nu \gamma \gamma$, which is the 
most relevant SM background to a signature with two acoplanar 
photons and large missing energy. Only very recently, dedicated 
calculations appeared in the literature, as motivated by the search of 
anomalous $\gamma\gamma + \rlap{$\, \, /$}E$ events at LEP. 
A complete diagrammatic calculation, 
using the helicity amplitude technique, 
supplemented with collinear SFs to account for ISR, was done in 
Ref.~\cite{mrenna97}. This paper confirmed a previous evaluation in 
Born approximation contained in Ref.~\cite{akkmm96}.   
Approximate predictions for the process of interest are obtained 
by the LEP collaborations by using the above quoted programs with 
QED ``dressing'' of the neutrino-pair cross section, 
namely PYTHIA (via the PS) and KORALZ (via the YFS method). Further, 
modern packages for the automatic calculation of Feynman 
amplitudes, such as GRACE~\cite{grace} and CompHEP~\cite{comphep}, 
are used by the experiments to calculate the cross section and 
generate events. Both packages implements collinear SFs for ISR, 
with an option for PS in GRACE.

Quite recently, an extensive comparison, at the level 
of total and differential cross sections, between all the 
available calculations has been performed in Ref.~\cite{bp97}.
This detailed comparison shows a (dis)agreement between 
independent calculations at 10-20\% level (or worse) 
and, more generally, 
an unsatisfactory situation about the software for the 
analysis of the data with acoplanar photons (see also last paper in 
Ref.~\cite{exp98}).

\section{Calculation of the lowest-order cross sections}

In order to approach the $O (1\%)$ theoretical precision and improve 
the predictions of the earlier version of the program NUNUGPV, 
the lowest-order matrix elements associated to the processes 
$e^+ e^- \to \nu \bar\nu n \gamma$ (with $n=1,2,3$) have been 
{\it exactly} calculated in the SM. 

\begin{figure}[ht]
\begin{center}
\epsfig{file=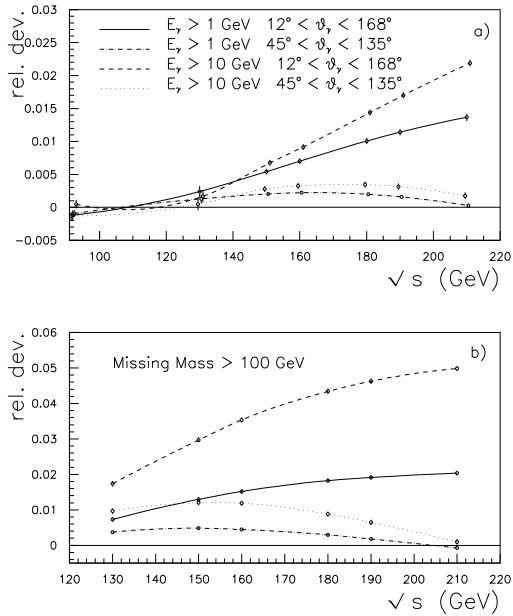,height=9truecm}
\end{center}
\caption{The relative deviation between the 
exact $e^+ e ^- \to \nu \bar \nu \gamma$ cross section 
and the approximated one, as obtained 
via eq.~(1) implemented in the earlier version of NUNUGPV. 
The differences 
are shown for the event selections given in Fig.~3a, 
with an additional 
cut on the missing mass in Fig.~3b.}
\label{fig:fig3}
\end{figure}

The matrix element for single-photon production has been computed 
by means of 
helicity amplitude techniques~\cite{helgp}, including the possibility 
of anomalous $\Delta k_{\gamma}$ and $\lambda_{\gamma}$ 
contributions to 
the $WW\gamma$ coupling. This offers the possibility of 
exploiting the LEP 
statistics relative to $\gamma + \rlap{$\, \, /$}E$ events 
in order to put constraints
on such anomalous terms. The diagrammatic calculation has 
been cross-checked by using the algorithm ALPHA~\cite{alpha} and found 
to be in perfect agreement. The three-body phase space 
has been generated 
recursively, by decomposing the phase-space element $d \Phi_3$ as 
follows
\begin{equation}
d \Phi_3 (P \to k,q_1,q_2) = d \Phi_2(P \to k, Q) 
(2 \pi)^3 d Q^2 d \Phi_2(Q \to q_1, q_2)  ,
\end{equation}
where $P = p_1 + p_2$ is the total incoming momentum, $Q = q_1 + q_2$ 
is the total momentum carried by the neutrinos and $k$ is the photon 
momentum. The independent kinematical variables are chosen to be: 
$E_\gamma$, $\cos\vartheta_\gamma$, $\phi_\gamma$, 
$\cos\theta_\nu^*$, $\phi_\nu^*$, where the photon variables 
are generated in the c.m. frame, while the neutrino ones in the 
reference frame where $\vec{q_1} + 
\vec{q_2} = 0$. Since the photon energy can be expressed 
in terms of the 
invariant mass $Q^2$ via the linear relation 
$E_\gamma = (s-Q^2)/2 \sqrt{s}$, where 
$s$ is the total c.m. energy, it is convenient to generate $Q^2$ 
in such a 
way to sample directly the leading matrix element configurations, 
that are  
due to the emission of a soft photon or a hard, $Z$ return one. 
This importance
sampling strategy is followed in the Monte Carlo integration 
in order to 
cure the variance of the matrix element in correspondence of the 
infrared and $Z$ return peaking behaviour. The photon angle 
is generated 
according to the weight function $p(\cos\vartheta_\gamma) 
\propto 1 / (1 - \beta^2 
\cos^2\vartheta_\gamma)$, with $\beta = \sqrt{1 - 4 m_e^2/ s}$, 
to take care of the collinear peaking.   

The exact treatment of the single-photon 
matrix element upgrades the released version of NUNUGPV, based on the 
photon spectrum of eq.~(\ref{eq:bs}) of Ref.~\cite{bbmmn88},
 to include 
previously neglected $W$-boson effects relative to 
contributions with at 
least three boson propagators. The size of such previously 
neglected effects 
is shown in Fig.~\ref{fig:fig3}, for typical event selections used 
by the LEP experiments~\cite{exp98,bus}. The calculation of the 
single-photon 
cross section obtained with the exact matrix element is compared with 
the cross section resulting from the integration of the 
photon spectrum 
of eq.~(\ref{eq:bs}). The relative difference between the 
two calculations 
is at a few per cent level, both without and with a cut 
on the missing mass, 
in agreement with the degree of approximation 
stated in Ref.~\cite{bbmmn88}. 
However, it should be noticed that an exact treatment 
of the lowest-order 
matrix element is actually mandatory at LEP2 if a theoretical accuracy
of the order of 1\% is aimed at. 

\begin{figure}[hbtp]
\begin{center}
\epsfig{file=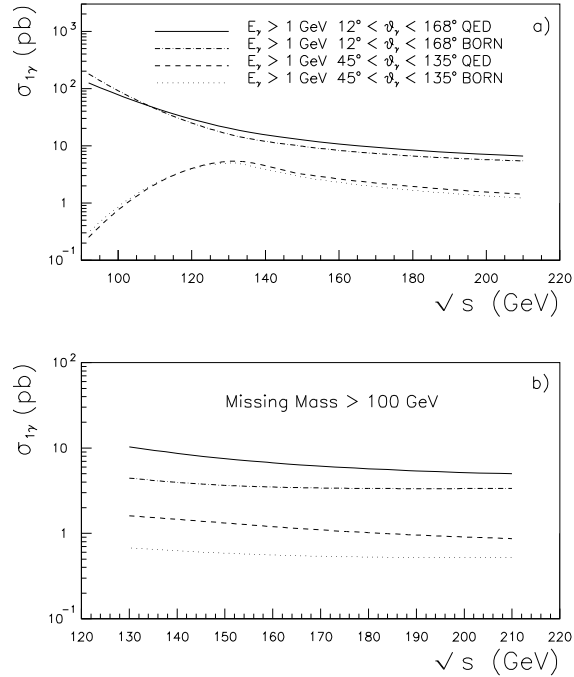,height=10.truecm}
\end{center}
\caption{The tree-level cross section for the process 
$e^+ e^- \to \nu \bar \nu \gamma$ 
as compared with the cross section with 
higher-order QED corrections, 
obtained by using collinear SFs. Two typical selection criteria, 
specified in Fig.~4a, are considered, including (Fig.~4a) and 
excluding, via 
a cut on the missing mass (Fig.~4b), the $Z$ radiative return.}
\label{fig:fig4}
\end{figure}

The already quoted algorithm ALPHA, that is conceived  
for the automatic computation of tree-level
multi-particle production amplitudes 
without any need of Feynman graphs expansion, has been employed for 
the calculation of the matrix elements with two and three 
photons in the 
final state. For the case $\nu \bar\nu \gamma\gamma\gamma$ the 
calculation here presented is the first one appearing in 
the literature.  
Without entering the details of the algorithm ALPHA, which is 
fully discussed in the literature~\cite{alpha}, it is worth noticing, 
for the aim of the present study, that the predictions of this 
automatic algorithm have been already compared with 
the diagrammatic results for the processes 
$e^+ e^- \to 4f$~\cite{yr96sm,yr96ww}, showing excellent agreement, 
and also successfully used to obtain original results for other 
reactions such as $\gamma\gamma \to 4f$~\cite{gg}, confirmed 
in Ref.~\cite{gg1}, $e^+ e^- \to 4f + \gamma$~\cite{4fg} and 
$e^+ e^- \to 6f$~\cite{6f}.
Concerning the treatment of phase space, the decomposition introduced 
for two photons in the final state reads as follows
\begin{equation}
d \Phi_4 (P \to k_1,k_2,q_1,q_2) = d \Phi_3(P \to k_1, k_2, Q) 
(2 \pi)^3 d Q^2 d \Phi_2(Q \to q_1, q_2)  ,
\end{equation}
where, as before, $Q = q_1 + q_2$ 
is the total momentum carried by the neutrinos and $k_1,k_2$ are 
the momenta of the two photons. 
The independent kinematical variables are chosen to be: 
$Q^2$, $E_{\gamma,1}$, $E_{\gamma,2}$, 
$\cos\vartheta_{\gamma,1}$, $\phi_{\gamma,1}$, $\phi_{\gamma,12}$, 
$\cos\theta_\nu^*$, $\phi_\nu^*$, where the photon variables 
are generated in the c.m. frame, while the neutrino ones in the 
reference frame where $\vec{q_1} + 
\vec{q_2} = 0$. As in the case of single-photon production, 
the energies of the photons are generated according 
to the soft and $Z$ 
return peaking structure, whereas one of the photon angles follow the 
collinear behaviour. 
A generalization of the above strategy is also followed 
for three photons in the final state. 

\begin{figure}[ht]
\begin{center}
\epsfig{file=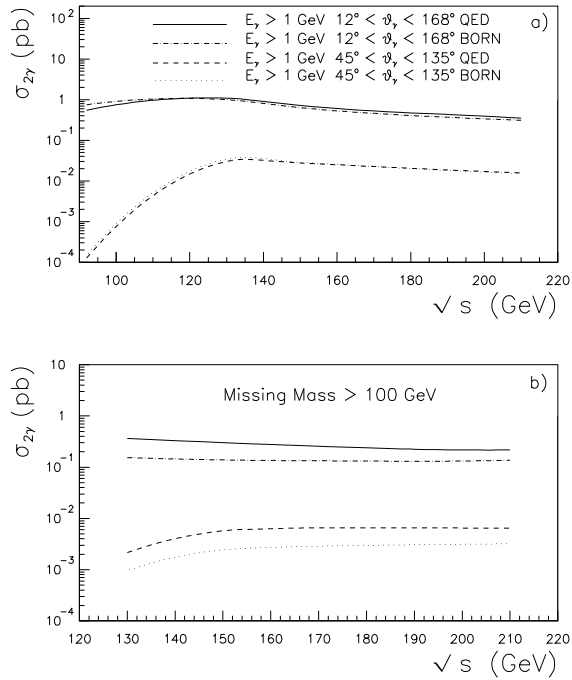,height=10.truecm}
\end{center}
\caption{The same as Fig.~4 for the 
process $e^+ e^- \to \nu \bar \nu \gamma \gamma$.}
\label{fig:fig5}
\end{figure}

Some numerical results for the 
lowest-order cross sections of the processes 
$e^+ e^- \to \nu \bar \nu \gamma$ and 
$e^+ e^- \to \nu \bar \nu \gamma \gamma$  
are shown, as functions of the c.m. energy in the LEP2 energy range,   
in Fig.~\ref{fig:fig4} and Fig.~\ref{fig:fig5}. The input parameters 
used throughout the present study are the same as 
in Ref.~\cite{mnpt95}.
By comparing Fig.~\ref{fig:fig4} and Fig.~\ref{fig:fig5}, 
it can be seen 
that the cross section for the signature $\nu \bar\nu \gamma\gamma$ 
is about a factor 10-100 smaller than the cross section for 
$\nu \bar\nu \gamma$, the reduction factor being strongly dependent, 
as expected, on the imposed photon cuts. 
A similar ratio is present in the 
LEP2 range for the $\nu \bar\nu \gamma \gamma \gamma$ cross section 
with respect to the 
one for the $\nu \bar\nu \gamma \gamma$ final state.

\section{Treatment of higher-order QED corrections}

Because the level of accuracy demanded to the calculations for 
single- and multi-photon production with missing energy should reach 
the $O(1\%)$ precision, the most important radiative corrections 
must be necessarily included. To this end, all the calculations 
known in the literature take into account the effect of ISR. 
As already 
discussed in Sect.~2, this goal is achieved by using standard
 algorithms 
for universal photonic corrections, such as the PS method, 
the YFS exponentiation and the SF approach. In particular, 
the SF approach, 
because of its simplicity, is certainly the most widely used 
algorithm, 
implemented in many generators of interest here, such 
as CompHEP, GRACE, MMM and NUNUGPV. More precisely, 
as discussed in Sect.~2, all the programs make use of SFs in 
strictly collinear 
approximation, while in NUNUGPV $p_t$-dependent 
SFs~\cite{nt89,nt89plb} are implemented to improve the
 treatment of ISR   
by including $p_t / p_L$ effects. Because of the 
presence of photons among 
the observed final-state products, the inclusion of the ISR requires 
a particular care. This caution is further motivated by the 
very large 
enhancement of the lowest-order cross sections as due to the ISR and 
clearly visible in Fig.~\ref{fig:fig4} and Fig.~\ref{fig:fig5}. 
As can be 
seen, the single- and double-photon cross section are significantly 
enhanced when considering typical event selections. The enhancement 
factor is about 1.3 when including the $Z$ return and about 2 when 
excluding it. As done in many practical 
applications~\cite{mnpt95,mmm,mrenna97}, the above results 
can be simply obtained by convoluting the 
hard-scattering cross section of interest 
with collinear SFs, according to the factorized formula
\begin{equation}
\sigma_{\mathrm{coll}} = \int d x_1 d x_2 \, 
D (x_1, s) D (x_2, s) \, d\sigma \, \Theta(cuts) .
\label{eq:sfcoll}
\end{equation}     
It allows to take into account the impact of 
higher-order QED corrections, 
due to photon emission before the hard-scattering 
reaction (pre-emission),   
at the LL level. 
It has been checked that the dominant 
effects of the ISR obtained according 
to eq.~(\ref{eq:sfcoll}) numerically agree, for different 
experimental set up, with those already 
known in the literature for one and two photons in the 
final state~\cite{mnpt95,mrenna97,bp97}.
Equation~(\ref{eq:sfcoll}) is a good approximation to QED
radiative corrections in the LEP1 energy regime at the LL 
level, since, with 
the standard selection criteria,  hard
``pre-emission'' photons imply the production of $\nu \bar\nu$
pairs off $Z$-resonance and are therefore inhibited. 
 On the contrary, going to LEP2 it can be   
easily realized that the implementation of the ISR as given 
by eq.~(\ref{eq:sfcoll}) is an approximation that 
clearly fails whenever 
the photonic degrees of freedom, of the pre-emission and 
hard-scattering
process, respectively, overlap in the same phase space region. 
Actually, 
because the collinear SFs can be seen as the result of 
an integration over 
the angular variables of the photon radiation, eq.~(\ref{eq:sfcoll}) 
does not take into account the correct statistical factor to
be included for the presence of identical particles in the 
final state.
Furthermore, if the pre-emission photon is detectable, the 
reconstruction of the event via eq.~(\ref{eq:sfcoll}) is only 
approximate 
and this might imply an additional inaccuracy.  
Therefore, one should expect that the implementation of the ISR 
as given 
by eq.~(\ref{eq:sfcoll}) leads to an overestimate of the higher-order 
QED corrections. This effect is clearly dependent on the 
photon(s) detection 
criteria and can be expected to be not negligible with respect to an 
$O(1\%)$ theoretical accuracy. An estimate of the effects due to the 
phase space overlapping of the IS pre-emission photons with 
the observed ones can be obtained by supplying the QED SFs with the
transverse degrees of freedom. Actually, the generation of the 
angular variables at the level of the ISR gives the 
possibility of rejecting in the event sample those pre-emission 
photons above the minimum detection angle, thus avoiding 
``overlapping effects''.  

According to such a procedure, the cross section with 
higher-order QED 
corrections can be calculated as follows (for the realistic 
data sample of at least one photon)
\begin{eqnarray}
\sigma^{1\gamma (\gamma)} &=& \int d x_1 d x_2 
d c_{\gamma}^{(1)} d c_{\gamma}^{(2)} \, 
\tilde{D} (x_1, c_{\gamma}^{(1)}; s) \tilde{D} 
(x_2, c_{\gamma}^{(2)}; s)  
\Theta(cuts) \nonumber \\
&& 
\quad \quad \qquad \quad \quad 
\qquad \times \left( d\sigma^{1\gamma} + d\sigma^{2\gamma} 
+ d\sigma^{3\gamma} + ... \right) ,
\label{eq:sfpt1}
\end{eqnarray}
where $\tilde{D} (x, c_{\gamma}; s)$~\cite{nunugpv} is 
a proper combination
of the collinear SF $D(x, s)$ with an 
angular factor inspired by the leading behaviour 
$1 / (p \cdot k)$. The latter is introduced to generate the 
angular variables 
of the pre-emission photons. According to eq.~(\ref{eq:sfpt1}), 
an ``equivalent'' photon is generated for each colliding lepton
and accepted as a higher-order ISR contribution  if:
\begin{itemize}
\item the energy of the equivalent photon is below the
 threshold for the
observed photon $E_{\gamma, min}$, for arbitrary angles;  or
\item the angle of the  equivalent photon is outside the angular 
acceptance for
the observed photons, for arbitrary energies. 
\end{itemize} 
\begin{figure}[ht]
\begin{center}
\epsfig{file=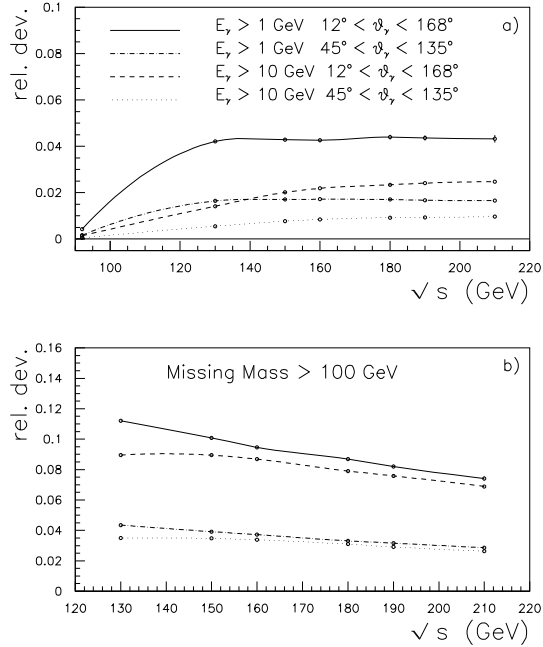,height=9.5truecm}
\end{center}
\caption{Contribution of ``overlapping effects'' 
(see the text for definition) to the cross section 
for the process $e^+ e ^- \to \nu \bar \nu \gamma (\gamma)$. 
In Fig.~2a the four lines correspond to the 
cuts on the photon energy and angle specified in the plot; 
Fig.~2b is the same 
as Fig.~2a, with an additional cut on the missing mass of the event.}
\label{fig:fig6}
\end{figure}
Within the angular acceptance of the seen photon(s), 
the cross section is evaluated by summing the exact matrix 
elements for 
the processes $e^+ e^- \to \nu \bar\nu n \gamma$, $n=1,2,3$ 
($d\sigma^{1\gamma}, d\sigma^{2\gamma}, d\sigma^{3\gamma}$). 
Notice that from the point of view of computing 
$\sigma^{1\gamma (\gamma)}$
the real contributions $d \sigma^{n\gamma}$, $n \ge 2$, represent the 
``hard'' radiative corrections to be matched with 
the universal soft+virtual ones
accounted for by the SFs. Therefore they are in principle
necessary at all orders. The truncation of hard radiative corrections 
at the level of  $d
\sigma^{3 \gamma}$ introduces a spurious infra-red 
sensitivity in radiative
corrections at the order $\alpha^4 \ln^4 (E / E_{\gamma,min})$ which,
from the practical point of view, is completely negligible at 
realistic 
$E_{\gamma,min}$.  Since the radiative corrections implemented by 
means of this
procedure are at the LL level, its  theoretical error is dominated by
missing truly $O (\alpha)$ corrections. 

In order to quantify the overestimate introduced by the 
collinear SFs in the calculation of the ISR via eq.~(\ref{eq:sfcoll}), 
the relative difference between eq.~(\ref{eq:sfcoll}) and 
eq.~(\ref{eq:sfpt1}) is shown in Fig.~\ref{fig:fig6}, for several 
photon detection criteria. As can be seen, at LEP1 the overlapping 
effects are contained within a few per mille, and 
therefore negligible 
on the scale of the
experimental accuracy. Going to LEP2 energies, the impact of 
``overlapping effects''~\footnote
{Notice that, as already remarked, by ``overlapping effects'' two 
effects are understood, which do occur simultaneously whenever
two or more photons are detected: in eq.~(\ref{eq:sfcoll}) 
$i)$ the statystical factor is
incorrectly accounted for, $ii)$ the reconstruction
of the event is approximate.} 
varies within 1-4\% when including the $Z$ return and is still larger, 
reaching 10\%, when imposing a cut on the missing mass, 
as usually done 
in realistic event selections. This effect is therefore important 
in the light of the aimed theoretical precision. It should be noticed,
however, that, where the difference reaches a ten per cent size, 
the overall effect of ISR is to enhance the tree-level cross section 
of a factor of two.
A qualitative explanation of the overlapping effects discussed 
above can be given as follows. 
The overestimate of radiative corrections
takes place when the pre-emission photon can reach the observability 
region for the detected photon. At LEP1, and with standard 
selection criteria, 
the emission of multiple detectable photons 
implies that neutrino production occurs off $Z$-resonance and 
therefore the overlapping effect is naturally suppressed 
by the dynamics. 
At LEP2, where this suppression is no longer active, 
the overlapping effects can become more sizeable, 
depending on the angular acceptance and minimum energy of the 
observed photons, and, more generally, in the presence of  
additional cuts on the four-momenta of the 
observed photons (as in the case of a cut on the missing mass). 

\begin{figure}[ht]
\begin{center}
\epsfig{file=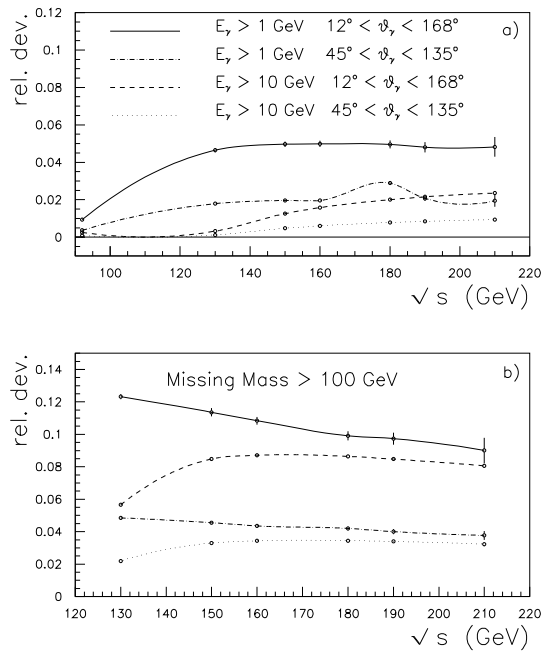,height=9.5truecm}
\end{center}
\caption{The same as Fig.~6 for the process
$e^+ e ^- \to \nu \bar \nu \gamma \gamma (\gamma) $.}
\label{fig:fig7}
\end{figure}

Analogously to eq.~(\ref{eq:sfpt1}), the QED corrected cross 
section for 
the signature of at least two-photons in the final state can be cast 
as follows 
\begin{eqnarray}
\sigma^{2\gamma (\gamma)} 
&=& \int d x_1 d x_2 
d c_{\gamma}^{(1)} d c_{\gamma}^{(2)} \,
\tilde{D} (x_1, c_{\gamma}^{(1)}; s) \tilde{D} 
(x_2, c_{\gamma}^{(2)}; s) 
\Theta(cuts) \nonumber \\ 
&& \quad \quad \qquad \quad \quad 
\qquad \times \left(d\sigma^{2\gamma} + d\sigma^{3\gamma} +... \right) .
\label{eq:sfpt2}
\end{eqnarray}
It is worth noticing that $d\sigma^{3\gamma}$ in
eq.~(\ref{eq:sfpt2}) plays the same role as $d\sigma^{2\gamma}$ in 
eq.~(\ref{eq:sfpt1}), and hence is a key ingredient when 
considering the
signature with at least two photons in the final state. 
Numerical results for such a signature are given in 
Fig.~\ref{fig:fig7}, showing the relative 
difference between eq.~(\ref{eq:sfcoll}) and 
eq.~(\ref{eq:sfpt2}). Also for the process 
$e^+ e ^- \to \nu \bar \nu \gamma \gamma (\gamma)$ 
the implementation of the ISR via collinear SFs can lead to an 
overestimate of the physical cross section of the 
order of several per cent.

\section{Distributions}
The results shown in the previous sections refer to integrated 
cross sections. However, in the search for NP effects, one is also 
interested in more exclusive distributions involving the 
photon(s) energy and angle. Therefore, it is worth studying 
how the theoretical improvements discussed above on tree-level 
matrix elements and higher-order QED corrections have an impact 
on photon distributions with respect to typical NP deviations.
\begin{figure}[ht]
\begin{center}
\epsfig{file=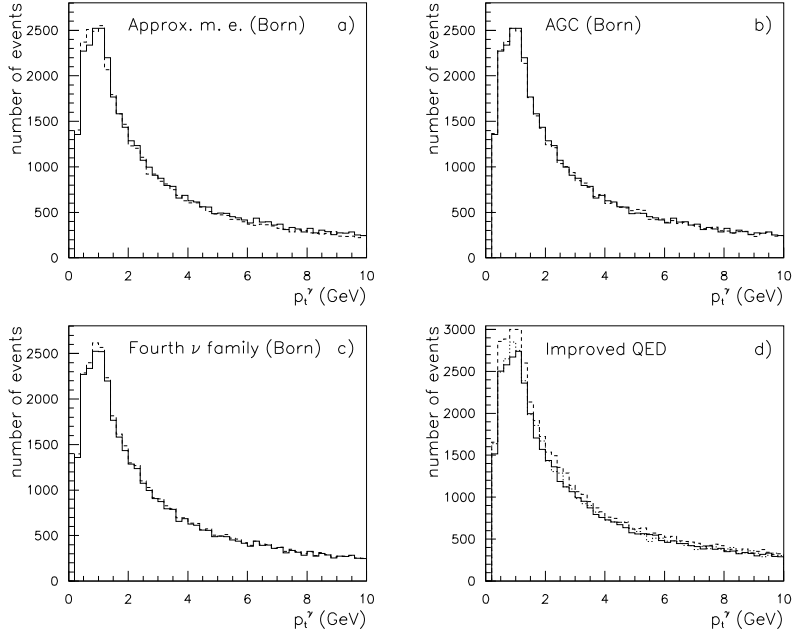,height=9.5truecm}
\end{center}
\caption{``Soft'' $p_t^\gamma$ distribution for 
 the leading photon of the process
$e^+ e ^- \to \nu \bar \nu \gamma (\gamma)$ at \protect{$\sqrt{s} = 
190$~GeV}. The cuts on the observed photons are 
$E_\gamma > 1$~GeV and $12^\circ < \vartheta_\gamma < 168^\circ$. 
Exact tree-level calculation (solid line) versus:
approximate analytic spectrum of Ref.~\cite{bbmmn88} 
(dashed line) (Fig.~8a), exact SM calculation plus anomalous 
couplings contribution (dashed line) (Fig.~8b), 
exact SM calculation plus 
massive ($m_\nu=50$ GeV)
fourth family neutrinos (dashed line) (Fig.~8c). 
In Fig.~8d ISR correction 
is implemented according to: eq.~(\ref{eq:sfpt1}) (solid line),
eq.~(\ref{eq:sfcoll}) (dashed line) and via SFs with $p_t/p_L$ 
effects as in the released version of NUNUGPV (dotted line).}
\label{fig:fig8}
\end{figure}

\begin{figure}[ht]
\begin{center}
\epsfig{file=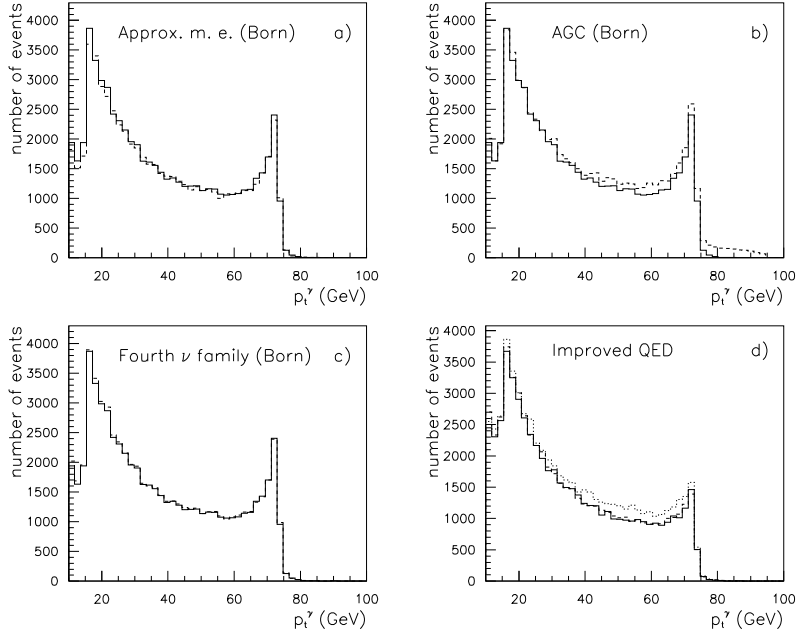,height=9.5truecm}
\end{center}
\caption{The same as Fig.~8, in the ``hard'' $p_t^\gamma$ region.}
\label{fig:fig9}
\end{figure}
\begin{figure}[ht]
\begin{center}
\epsfig{file=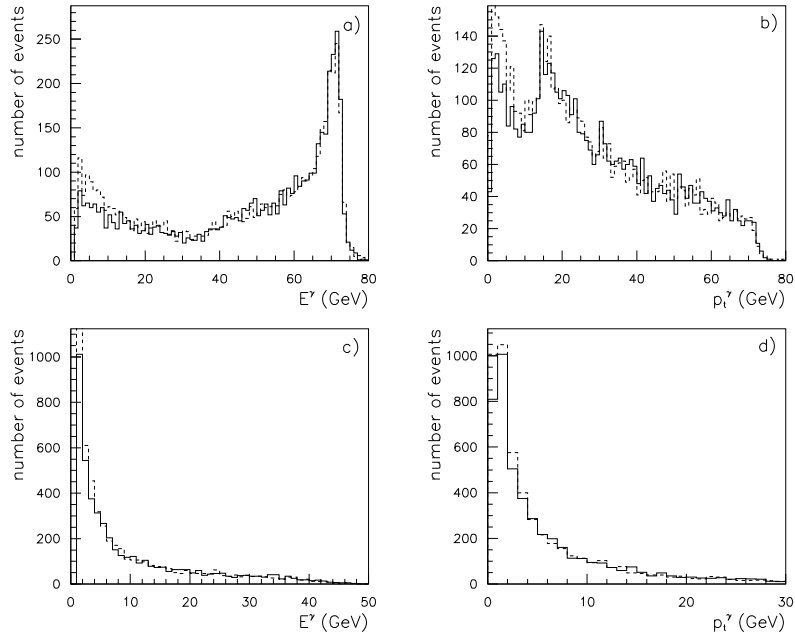,height=9.5truecm}
\end{center}
\caption{$E_\gamma$ and $p_t^\gamma$ distributions 
for the leading (Figs.~10a-10b) and next-to-leading photon 
(Figs.~10c-10d) of the process
$e^+ e ^- \to \nu \bar \nu \gamma \gamma (\gamma)$ at 
\protect{$\sqrt{s} = 190$~GeV}. The cuts on the observed photons are 
$E_\gamma > 1$~GeV and $12^\circ < \vartheta_\gamma < 168^\circ$. 
ISR correction is implemented according to: eq.~(\ref{eq:sfpt2}) 
(solid line) 
and eq.~(\ref{eq:sfcoll}) (dashed line).}
\label{fig:fig10}
\end{figure}

 
To this aim a few results obtained by using the improved version 
of NUNUGPV as an event generator are illustrated 
in Figs.~\ref{fig:fig8}-\ref{fig:fig10}. Figure~\ref{fig:fig8} and 
Fig.~\ref{fig:fig9} refer to the signature 
$e^+ e ^- \to \nu \bar \nu \gamma (\gamma) $, 
while Fig.~\ref{fig:fig9} is relative to the process with at least 
two photons, i.e. $e^+ e ^- \to \nu \bar \nu \gamma \gamma (\gamma)$. 
For the sake of comparison, the histograms shown 
in Figs.~\ref{fig:fig8}-\ref{fig:fig10} are normalized 
to the same luminosity. A typical LEP2 energy of 
$\sqrt{s} = 190$~GeV is considered, with the cuts $E_\gamma > 1$~GeV 
and $12^\circ < \vartheta_\gamma < 168^\circ$ 
for the observed photons.

As an example, the $p_t^\gamma$ distribution of the 
most energetic photon    
(leading photon) is shown  in Fig.~\ref{fig:fig8}  for the ``soft'' 
region and 
in Fig.~\ref{fig:fig9} for the ``hard'' one. 
The first three plots (Figs.~8a-8c, Figs.~9a-9c) 
do not take into account the 
contribution of ISR (Born approximation), while the last one (Fig.~8d,
Fig.~9d) shows QED corrected distributions according to different 
realizations. 
The aim is to show how typical NP 
effects may compete with an improved calculation of the 
SM background, both at the level of tree-level 
matrix elements and higher-order QED corrections, 
previously discussed. 
Actually, 
in the first three plots the solid line is the lowest-order prediction 
obtained by means of the exact single-photon matrix element 
as compared 
with the following results (dashed histograms)
\begin{enumerate}

\item approximate lowest-order photon spectrum of eq.~(\ref{eq:bs}), 
implemented 
in the released version of NUNUGPV (Fig.~8a, Fig.~9a);

\item exact single-photon matrix element with anomalous 
$WW\gamma$ coupling, 
corresponding to the parameters choice 
$\Delta k_\gamma = \lambda_\gamma = 5$ (Fig.~8b, Fig.~9b);

\item exact single-photon matrix element with additional 
contribution due 
to the production of a pair of fourth-generation neutrinos, with 
standard couplings 
and mass of 50~GeV (Fig.~8c, Fig.~9c).

\end{enumerate} 
As can be seen, an exact treatment of the lowest-order matrix element 
is mandatory 
in order to obtain fully reliable exclusive photon distributions and 
to avoid loss of sensitivity in the search of (small) NP 
deviations. Also a careful formulation of ISR is necessarily required 
for an appropriate
simulation of photon distributions, as shown 
in Fig.~\ref{fig:fig8}d and Fig.~\ref{fig:fig9}d. 
In these plots the three following 
different implementations of higher-order QED corrections 
are compared: 
improved treatment of ISR via eq.~(\ref{eq:sfpt1}) (solid histogram),
simulation of ISR via collinear SFs as given by eq.~(\ref{eq:bs}) 
(dashed histogram), 
treatment of ISR via SFs with $p_t/p_L$ effects as in 
the released version 
of NUNUGPV (dotted histogram). The differences between the three 
implementations 
of ISR are certainly comparable or larger than the deviations 
introduced by NP, 
clearly illustrating the need of a proper treatment of 
pre-emission QED effects. 
In particular, the collinear SFs lead, as expected, to an 
overestimate of 
events in the soft-photon region, while the implementation 
of ISR via SFs 
with $p_t/p_L$ effects, as in the 
released version of NUNUGPV, is responsible for an excess of events 
in the large $p_t^\gamma$ region, as 
a consequence of the approximate treatment of $p_t$ 
contributions outside the collinear approximation. 

A comparison between photon distributions obtained by means of 
different simulations of 
ISR is also shown in Fig.~\ref{fig:fig10} for the 
$e^+ e ^- \to \nu \bar \nu \gamma \gamma (\gamma)$ events. 
The implementation of 
ISR via collinear SFs as in eq.~(\ref{eq:sfcoll}) (dotted histograms) 
is compared with 
the improved formulation of eq.~(\ref{eq:sfpt2}) (solid histograms). 
The energy 
and $p_t$ distributions of the leading (Figs.~10a-10b) and of 
the next-to-leading (Figs.~10c-10d) 
photon are considered. It can be seen that the implementation 
of ISR via collinear 
SFs, as done in practice in many computational tools 
for $\nu \bar \nu \gamma \gamma$ 
final state, lead to an excess of events in the soft-photon region, 
as already noticed 
for the $\nu \bar \nu \gamma$ signature. Since the 
soft-photon region is of 
primary interest for the search of NP effects, 
a careful treatment of ISR, as 
given by eqs.~(\ref{eq:sfpt1})-(\ref{eq:sfpt2}), 
is essential to obtain 
actually precise predictions for the SM irreducible background.
 
\section{Concluding remarks}

The search for new physics in single- and multi-photon
 final states with 
large missing energy at LEP and future $e^+ e^-$ colliders requires 
the best knowledge of the SM irreducible background. To this aim, 
precise calculations (and related computational tools) for 
the processes $e^+ e^- \to \nu \bar\nu n\gamma$ are
 presently demanded. 
Towards such a direction, the tree-level matrix elements for the 
SM processes 
with neutrino pairs and up to three photons in the final state have 
been 
calculated without any approximation. The exact treatment of the
 lowest-order 
transition amplitudes has been seen to be actually necessary in view 
of an expected  precision at the 1\% level. At this accuracy 
level, also a careful treatment of the (large) effect of 
the higher-order 
corrections introduced by ISR is unavoidable. Indeed, it has been   
shown that the usual implementation of ISR via collinear SFs, 
which is a good approximation at LEP1 energies and with the 
usual selection 
criteria, can lead to an overestimate of the physical cross section 
much larger than 1\%. 
In particular, in the presence of a missing mass cut, the 
ISR overestimate of the integrated cross section reaches the ten 
percent size.
Furthermore, significant effects, whenever 
compared with typical NP deviations, have been
shown to be present also in the photon distributions 
and should therefore 
carefully considered in a sensible experimental analysis. 
The remaining uncertainty in the present study is left to 
the yet unknown exact $O (\alpha)$ electroweak corrections to the 
process $e^+ e^- \to \nu \bar\nu \gamma$. Such a 
complete calculation should be actually desirable  
to reach a theoretical error not exceeding the 1\% level.

As a result of the present study, an improved version of the 
event generator NUNUGPV is by now available. It includes 
the exact SM matrix
elements for $\nu \bar\nu n\gamma$ production, with $n=1,2,3$, 
and a careful treatment of the ISR 
including $p_t/p_L$ effects. For single-photon production the 
possibility of studying the effects of anomalous couplings 
is included. 
Predictions for the production of 
a pair of hypothetic massive neutrinos 
can be also obtained. A sample of numerical results showing the 
potentials of the 
new version of NUNUGPV has been shown, with particular emphasis 
on the impact of the discussed improvements on the 
integrated cross sections 
and more exclusive distributions.
The program can be used for a full analysis of single- and 
multi-photon events with missing energy at LEP2 and future high-energy 
$e^+ e^-$ colliders.

\vskip 48pt\noindent
{\large\bf Acknowledgements}\\
\vskip 4pt\noindent
We wish to thank J.~Busenitz, P.~Checchia, C.~Matteuzzi, B.~Mele and 
G.W.~Wilson 
for valuable information, useful discussions and interest in our work. 
\vfill\eject
\bibliography{multig}

\end{document}